\documentclass[12pt]{iopart}

\usepackage{graphicx}
\usepackage{amssymb}
\usepackage{dsfont}
\usepackage{bm}
\usepackage{mathrsfs}
\usepackage{times}

\begin{document}
\title{No quantum friction between uniformly moving plates}

\author{T G Philbin and U Leonhardt}

\address{School of Physics and Astronomy, University of St Andrews, North Haugh, St Andrews. Fife KY16~9SS, Scotland}
\ead{tgp3@st-andrews.ac.uk}

\begin{abstract}
The Casimir forces between two plates moving parallel to each other at arbitrary constant speed are found by calculating the vacuum electromagnetic stress tensor. The perpendicular force between the plates is modified by the motion but there is no lateral force on the plates. Electromagnetic vacuum fluctuations do not therefore give rise to ``quantum friction" in this case, contrary to previous assertions. The result shows that the Casimir--Polder force on a particle moving at constant speed parallel to a plate also has no lateral component.
\end{abstract}

\section{Introduction}
Over half a century ago, the vacuum force between two parallel dielectric plates was calculated by Lifshitz~\cite{lif55,LL}. Casimir's simple formula~\cite{cas48} for the vacuum force between two perfect mirrors then emerges in the limit of infinite permittivity. A generalization of the problem to where one of the plates moves at a constant speed parallel to the other might be expected to be straightforward, but in fact authors have obtained conflicting answers for the vacuum forces in this case~\cite{teo78,sch81,lev89,pol90,mkr95,bar96,pen97,per98,vol99}. Interest in this problem has attached particularly to the lateral component of the Casimir force, which would give information on the extent of a quantum vacuum contribution to friction between bodies~\cite{teo78,sch81,lev89,pol90,mkr95,bar96,pen97,per98,vol99,kar99,kri02,dav05}. In this paper we use Lifshitz theory~\cite{lif55,dzy61,LL} to find the zero-temperature Casimir forces between the plates for arbitrary velocity. Despite a general acceptance in the literature on this problem~\cite{teo78,sch81,lev89,pol90,mkr95,bar96,pen97,per98,vol99} that it provides an example of quantum friction, the exact solution, presented here for the first time, shows that the motion does not in fact induce a lateral Casimir force.

\begin{figure}[t]
\begin{center}
\includegraphics[width=12.0pc]{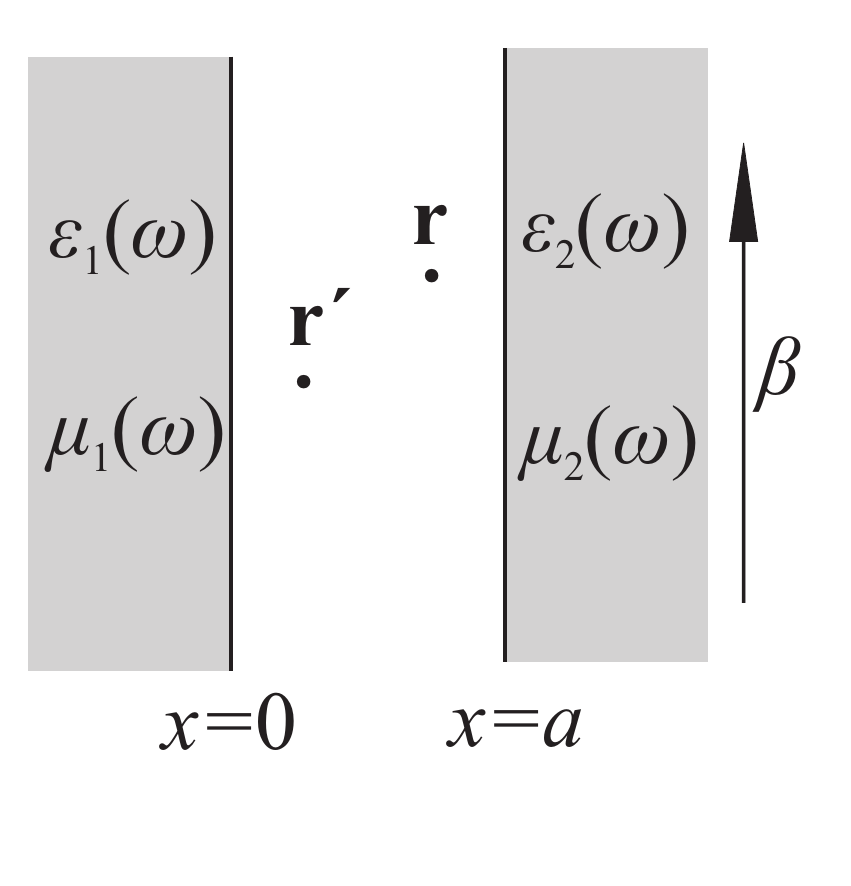}
\vspace*{-12mm}
\end{center}
\caption{Two plates with general electric permitivities and magnetic permeabilities lie in the $yz$-plane with constant separation $a$. The second plate moves in the postive $y$-direction with speed $\beta c$, where $c$ is the speed of light. We compute the Casimir forces on Plate~1. \label{fig1}}
\end{figure}

The geometry and notation of the problem are set out in Figure~\ref{fig1} and caption. The plates have arbitrary permittivities and permeabilities and are taken to have a separation $a$ in the $x$-direction. Plate~2 moves in the positive $y$-direction and we use $\beta$ to denote its speed in units of the speed of light $c$, i.e.\ the speed is $\beta c$.
It is clear that the arrangement in Figure~\ref{fig1} represents one of the most obvious and basic problems in the theory of Casimir forces and the relatively limited attention it has received is somewhat surprising (a dozen papers or so, over the past thirty years). The practical significance of this problem has also increased with the emerging importance of Casimir forces in micro- and nano-engineering~\cite{nature,cap07}. Indeed, the problem in question is the simplest possible model of a nanomachine. 

Before entering into technical details it is worth considering what one might reasonably expect for the Casimir forces in this case. Casimir forces are caused by the vacuum zero-point modes of the electromagnetic field~\cite{mil94}. If the plates in Figure~\ref{fig1} are perfect mirrors then the zero-point modes do not penetrate the materials and the motion of Plate~2 can therefore have no effect on the vacuum forces: there is an attractive force between the mirrors given by Casimir's formula~\cite{cas48} with no lateral component. For realistic materials, however, the zero-point modes will penetrate the moving plate and the motion should therefore affect the Casimir force; the perpendicular force is indeed altered by the motion, but the intriguing question is whether the vacuum modes are influenced in a way that produces a lateral force on the plates. At first glance a lateral force may seem plausible: the quantum vacuum can sometimes be usefully thought of as a kind of fluid~\cite{dav05}, and one could imagine this fluid exerting a drag on the moving plate, with the reciprocal drag of the vacuum exerting a lateral force on the non-moving plate. The quantum vacuum is a very strange kind of fluid, however: uniform motion relative to the quantum vacuum has no meaning because of the Lorentz invariance of quantum field theory, so an isolated moving body only experiences a quantum-vacuum ``viscosity'' when it accelerates, not when it moves at a constant velocity~\cite{bir84}. Whether the presence of the non-moving plate in Figure~1 is enough to make the quantum vacuum become viscous to the moving plate (and to the non-moving one) is far from clear. The following consideration suggests the absence of a lateral force. A moving medium is equivalent to a particular non-moving bi-anisotropic medium~\cite{LLcm,GREE}; for the present argument the bi-anisotropy can be very small, so it is hard to see any reason why such a material could not be constructed in principle. It would be very strange if a bi-anisotropic medium could be used to induce a unidirectional lateral Casimir force as this would seem to allow the extraction of unlimited energy from the quantum vacuum. If anisotropy is introduced into both plates then lateral Casimir forces are certainly possible, but these are rather torques that act to orient the plates to an equilibrium position~\cite{bar78,mun05,phi08,ros08}. 

Almost all of the treatments of this problem~\cite{teo78,sch81,lev89,pol90,mkr95,bar96,pen97,per98,vol99} have made use of various approximations, and there is disagreement on the magnitude of the Casimir forces. We make particular mention of the paper by Barton~\cite{bar96}, although the concern in that analysis was almost exclusively with the case where the moving plate in Figure~1 is accelerating rather than moving at a constant speed. The approximations used in~\cite{bar96} included ignoring the dispersion and absorption of the plates and considering only the lowest-order effects of the motion. The lateral force on the plates was found to vanish in the case of constant velocity. Barton noted~\cite{bar96} that this result contradicted previous treatments of the constant-velocity case in~\cite{teo78,sch81,lev89} (which also contradict each other) but because of the different and severely limited methods used to model the materials he drew no conclusions as to the correct result for realistic materials. All other treatments of the problem with which we are concerned conclude that there is a lateral force on the plates. The previous claims~\cite{vol99} (see also~\cite{vol08}) of an exact solution were based on Rytov's theory of electromagnetic fluctuations, which utilizes the classical Green tensor for the problem to calculate the forces. Since our results are also derived from the classical Green tensor it is here that our work makes direct contact with the existing literature. As we show in Section~\ref{Stress},  the Green tensor was not calculated correctly in~\cite{vol99}, even approximately (essentially the same error was also made in~\cite{pen97}). In fact, the classical Green tensor represents the main computational challenge in this problem and it is presented here for the first time.  

In Section~\ref{theory} we show how the vacuum electromagnetic stress tensor is related to the classical Green tensor; this is similar to the usual Lifshitz theory~\cite{lif55,dzy61,LL} except that we treat only zero temperature. The exact Green tensor for the problem is derived in Section~\ref{Green}; this is achieved by a novel method which utilizes physical reasoning in place of considerable algebraic labour. Section~\ref{Stress} presents the Casimir force on the plates, and in our Conclusions we discuss the implications for the Casimir--Polder force between a particle and a plate.

\section{Lifshitz theory at zero temperature} \label{theory}
Casimir forces are given by the vacuum expectation value of the electromagnetic stress tensor~\cite{jac}:
\begin{equation} \label{stress}
\bm{\sigma}=\varepsilon_0\langle \mathbf{\hat{E}}\otimes\mathbf{\hat{E}}\rangle+\mu_0^{-1}\langle \mathbf{\hat{B}}\otimes\mathbf{\hat{B}}\rangle-\frac{1}{2}\mathds{1}(\varepsilon_0\langle \hat{E}^2\rangle+\mu_0^{-1}\langle \hat{B}^2\rangle).
\end{equation}
The methodology of Lifshitz theory is to compute the expectation values in the vacuum stress (\ref{stress}) using the retarded Green tensor of the classical vector potential $\mathbf{A}$ in a gauge in which the scalar potential set to zero~\cite{LL}. In this gauge the electric and magnetic fields are given by
\begin{equation}  \label{EBdef}
\mathbf{E}=-\partial_t\mathbf{A}, \qquad \mathbf{B}=\nabla\times\mathbf{A},
\end{equation}
and the vector-potential wave equation is
\begin{equation} \label{Awave}
\left(\nabla\times\nabla\times+\frac{1}{c^2}\partial^2_t\right)\mathbf{A}=\mu_0\mathbf{j},
\end{equation}
where $\mathbf{j}$ is the current density of any sources. The retarded Green tensor for the vector potential satisfies
\begin{equation}  \label{greent}
\left(\nabla\times\nabla\times+\frac{1}{c^2}\partial^2_t\right)\mathbf{G}(\mathbf{r},t;\mathbf{r'},t')=\mathds{1}\delta(\mathbf{r}-\mathbf{r'})\delta(t-t'),
\end{equation}
and relates the vector potential at $\mathbf{r},t$ to the current density $\mathbf{j}$ at $\mathbf{r'},t'$, where $t'<t$:
\begin{equation}  \label{AGj}
\mathbf{A}(\mathbf{r},t)=\mu_0\int_{-\infty}^t\rmd t'\int\rmd^3\mathbf{r'}\,\mathbf{G}(\mathbf{r},t;\mathbf{r'},t')\cdot\mathbf{j}(\mathbf{r'},t').
\end{equation}
The retarded boundary condition means that $\mathbf{G}(\mathbf{r},t;\mathbf{r'},t')=0$ for $t'>t$ and in frequency space the Green tensor is defined by
\begin{eqnarray}
\mathbf{G}(\mathbf{r},\mathbf{r'},\omega)=\int^\infty_{0}\rmd t \,\mathbf{G}(\mathbf{r},t,\mathbf{r'},0)\,\rme^{\rmi \omega t},  \label{freqG} \\[5pt]
\left(\nabla\times\nabla\times-\frac{\omega^2}{c^2}\right)\mathbf{G}(\mathbf{r},\mathbf{r'},\omega)=\mathds{1}\delta(\mathbf{r}-\mathbf{r'}). \label{green}
\end{eqnarray}
Since a real current will produce a real vector potential, (\ref{AGj}) shows the Green tensor is real and in frequency space this gives the property
\begin{equation} \label{syms}
\mathbf{G}(\mathbf{r},\mathbf{r'},-\omega^*)=\mathbf{G}^*(\mathbf{r},\mathbf{r'},\omega).
\end{equation}
The similar property $\varepsilon(-\omega^*)=\varepsilon^*(\omega)$ of the dielectric permittivity is derived in an identical manner~\cite{LLcm}.

In (\ref{greent}) we have written the equation for the Green tensor in the vacuum between the plates; the Green tensor inside the plates will not be required to impose the boundary conditions on (\ref{green}) since we reduce the problem to a consideration of plane waves, the boundary conditions for which are imposed by their standard reflection and transmission coefficients~\cite{jac}. Equation (\ref{stress}) is also valid only in the vacuum between the plates. The net force on the plates is a result of the electromagnetic stress tensor in the plates as well in the vacuum but it turns out that the relevant stresses in the plates are zero, as is discussed in detail in Section~\ref{Stress} and \ref{A2}. 

At finite temperature it can be shown~\cite{LL}, through use of the fluctuation-dissipation theorem and an assumption of time-symmetric boundary conditions, that  the retarded Green tensor  has a simple relation to a correlation function for the vector potential. Clearly this derivation does not apply to the problem considered here. For zero-temperature problems there is in fact no need to invoke the fluctuation-dissipation theorem and there is a much simpler proof~\cite{leo} of the required relations.

The vector potential operator $\mathbf{\hat{A}}(\mathbf{r},t)$ is given by~\cite{lou}
\begin{equation}    \label{As}
\mathbf{\hat{A}}(\mathbf{r},t)=\sum_{\mathbf{k},\sigma}\left[\mathbf{A}_{\mathbf{k},\sigma}(\mathbf{r})\rme^{-\rmi \omega_k t}\hat{a}_{\mathbf{k},\sigma}+\mathbf{A}^*_{\mathbf{k},\sigma}(\mathbf{r})\rme^{\rmi \omega_k t}\hat{a}_{\mathbf{k},\sigma}^\dagger\right],
\end{equation}
where $\mathbf{A}_{\mathbf{k},\sigma}(\mathbf{r})$, $\mathbf{A}^*_{\mathbf{k},\sigma}(\mathbf{r})$ are a complete set of modes and $\sigma$ labels two linearly independent polarizations. We use the standard technique~\cite{lou} of a finite quantization cavity, the volume of which can eventually be increased without limit. The modes are solutions of the homogeneous equation
\[
\left(\nabla\times\nabla\times-\frac{\omega_k^2}{c^2}\right)\mathbf{A}_{\mathbf{k},\sigma}(\mathbf{r})=0
\]
and have the normalization
\begin{equation} \label{norms}
\fl  \int \mathbf{A}^*_{\mathbf{k},\sigma}(\mathbf{r})\cdot \mathbf{A}_{\mathbf{k'},\sigma'}(\mathbf{r})\,\rmd^3x=\frac{\hbar}{2\varepsilon_0\omega_k}\mathds{1}\delta_{kk'}\delta_{\sigma\sigma'}, \qquad
\int \mathbf{A}_{\mathbf{k},\sigma}(\mathbf{r})\cdot \mathbf{A}_{\mathbf{k'},\sigma'}(\mathbf{r})\,\rmd^3x=0.
\end{equation}

The Green tensor (\ref{freqG})--(\ref{green}) can be expanded in terms of the eigenfunctions (\ref{As}) by a standard procedure (see~\cite{jac} for example). A (positive frequency) Green tensor $\mathbf{G}(\mathbf{r},\mathbf{r'},\omega)$ with the same boundary conditions as the modes  $\mathbf{A}_{\mathbf{k},\sigma}(\mathbf{r})$ can be expanded in terms of these modes. We can therefore write
\begin{equation} \label{expans}
\mathbf{G}(\mathbf{r},\mathbf{r'},\omega)=\sum_{\mathbf{k},\sigma}\mathbf{A}_{\mathbf{k},\sigma}(\mathbf{r})\otimes \mathbf{c}_{\mathbf{k},\sigma}(\mathbf{r'}).
\end{equation}
Acting on (\ref{expans}) with the operator on the left-hand side of (\ref{green}) we obtain
\[
\sum_{\mathbf{k},\sigma}\frac{\omega_k^2-\omega^2}{c^2}\mathbf{A}_{\mathbf{k},\sigma}\otimes \mathbf{c}_{\mathbf{k},\sigma}=\mathds{1}\delta(\mathbf{r}-\mathbf{r'}).
\]
A dot product with $\mathbf{A}^*_{\mathbf{k},\sigma}(\mathbf{r'})$ and use of (\ref{norms}) then determine $\mathbf{c}_{\mathbf{k},\sigma}$ and we find the standard expansion~\cite{jac}
\begin{equation} \label{expan2s}
\mathbf{G}(\mathbf{r},\mathbf{r'},\omega)=\frac{2\varepsilon_0c^2}{\hbar}\sum_{\mathbf{k},\sigma}\frac{\omega_k}{\omega_k^2-\omega^2}\mathbf{A}_{\mathbf{k},\sigma}(\mathbf{r})\otimes \mathbf{A}^*_{\mathbf{k},\sigma}(\mathbf{r'}).
\end{equation}

Consider the equal-time vacuum correlation function for the electric field: $\langle\mathbf{\hat{E}}(\mathbf{r},t)\otimes\mathbf{\hat{E}}(\mathbf{r'},t)\rangle$. Using  (\ref{As}) and the operator version of (\ref{EBdef}) we find
\begin{equation} \label{Ecors}
\langle\mathbf{\hat{E}}(\mathbf{r},t)\otimes\mathbf{\hat{E}}(\mathbf{r'},t)\rangle=\sum_{\mathbf{k},\sigma} \omega_k^2\,\mathbf{A}_{\mathbf{k},\sigma}(\mathbf{r})\otimes \mathbf{A}^*_{\mathbf{k},\sigma}(\mathbf{r'}).
\end{equation}
The right-hand side of (\ref{Ecors}) is proportional to a contour integration of the Green tensor (\ref{expan2s}) in the complex $\omega$ plane. 
\begin{figure}
\includegraphics[width=38.0pc]{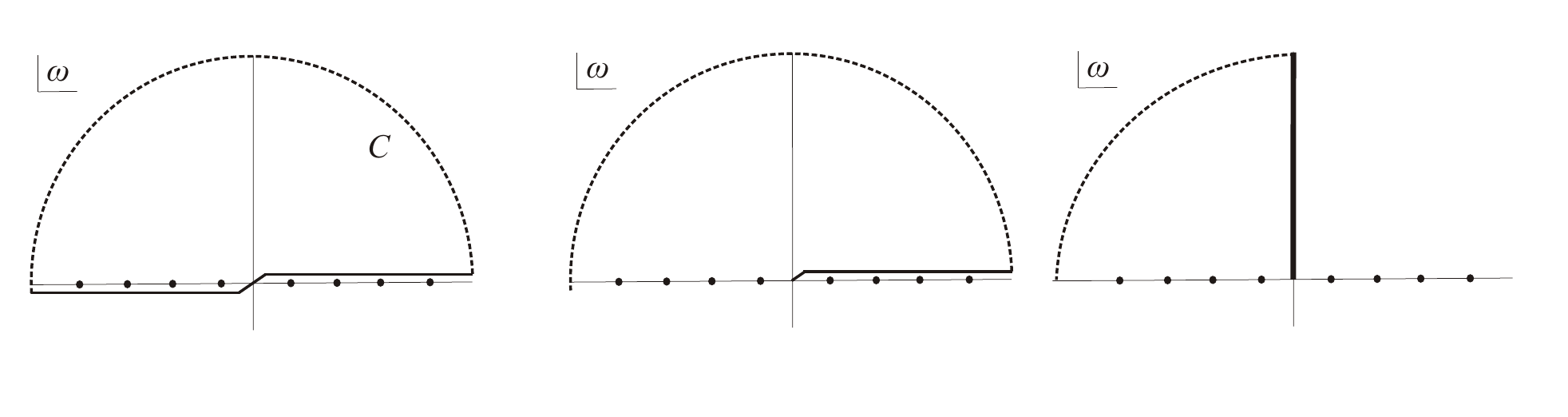}
\vspace*{-15mm}
\caption{Contours in the complex $\omega$-plane. The left picture shows the contour $C$. In the centre picture the integration below the negative real axis has been replaced by an additional integration above the positive real axis, as described in the text. In the right picture the contour above the poles on the positive real axis has been rotated to the imaginary axis.   \label{fig2}}
\end{figure}
The denominator $\omega_k^2-\omega^2=(\omega_k-\omega)(\omega_k+\omega)$ gives poles in the complex $\omega$ plane at all $\omega_k$ and $-\omega_k$. Consider a contour $C$ that lies just below the poles at $-\omega_k$ on the negative real axis and just above the poles $-\omega_k$ on the positive real axis and which is then closed in the upper half complex plane (see Figure~\ref{fig2}). From (\ref{freqG}), $\mathbf{G}(\mathbf{r},\mathbf{r'},\omega)$ is analytic in the upper half plane and vanishes for $\mathrm{Im}(\omega)\rightarrow\infty$. An integration of (\ref{expan2s}) over the contour $C$ therefore has contributions from the $-\omega_k$ poles only. In this manner we find
\begin{equation} \label{Cs0}
\langle\mathbf{\hat{E}}(\mathbf{r},t)\otimes\mathbf{\hat{E}}(\mathbf{r'},t)\rangle=\frac{\hbar}{2\varepsilon_0c^2}\frac{1}{\rmi \pi}\int_{C}\mathbf{G}(\mathbf{r},\mathbf{r'},\omega)\omega^2 \rmd\omega.
\end{equation}
In order to compute the stress tensor (\ref{stress}) we require the correlation function (\ref{Cs0}) at $\mathbf{r}=\mathbf{r'}$. The electric field operators commute with each other at space-like separated events, but the limit $\mathbf{r}\rightarrow\mathbf{r'}$ is singular and must be regularized (see below). To avoid ambiguities in the operator ordering we follow the usual prescription of quantum mechanics and define the correlation function at $\mathbf{r}=\mathbf{r'}$ by
\[
\langle\hat{E}_i(\mathbf{r},t)\otimes\hat{E}_j(\mathbf{r},t)\rangle=\lim_{\mathbf{r}\rightarrow\mathbf{r'}}\frac{1}{2}\left(\langle\hat{E}_i(\mathbf{r},t)\otimes\hat{E}_j(\mathbf{r'},t)\rangle+\langle\hat{E}_j(\mathbf{r'},t)\otimes\hat{E}_i(\mathbf{r},t)\rangle\right).
\]
We then obtain from (\ref{Cs0})
\begin{equation} \label{Cs}
\langle\mathbf{\hat{E}}(\mathbf{r},t)\otimes\mathbf{\hat{E}}(\mathbf{r},t)\rangle=\frac{\hbar}{2\varepsilon_0c^2}\frac{1}{\rmi \pi}\int_{C}\mathbf{G}^S(\mathbf{r},\mathbf{r},\omega)\omega^2 \rmd\omega,
\end{equation}
where $\mathbf{G}^S$ is the symmetrized Green tensor, given by
\begin{equation} \label{GS}
G^S_{ij}(\mathbf{r},\mathbf{r'},\omega)=\frac{1}{2}\left[G_{ij}(\mathbf{r},\mathbf{r'},\omega)+G_{ji}(\mathbf{r'},\mathbf{r},\omega)\right],
\end{equation}

Since $\mathbf{G}(\mathbf{r},\mathbf{r'},\omega)$ vanishes for $\mathrm{Im}(\omega)\rightarrow\infty$ (see (\ref{freqG})), the upper part of the semicircle in  Figure~\ref{fig2} does not contribute to the integral (\ref{Cs}). Moreover the contribution from large real $\omega$ does not contribute either due to the regularization procedure required for Casimir calculations in which the bare Green tensor in the absence of the plates is subtracted from $\mathbf{G}(\mathbf{r},\mathbf{r'},\omega)$~\cite{LL}; this subtraction gives zero at large frequencies where the plates are transparent and macroscopic electromagnetism breaks down~\cite{LL}. Hence there is no contribution from the entire semicircle in Figure~\ref{fig2}.

Using (\ref{syms}), the integration in (\ref{Cs}) over the part of the contour $C$ that lies along the negative real $\omega$-axis can be replaced by an integration along the positive real $\omega$-axis (see Figure~\ref{fig2}). Since the contour lies above the poles on the positive real axis, it can be pushed up to the imaginary axis (see Figure~\ref{fig2}). Eqn.~(\ref{syms}) shows that the Green tensor is real for imaginary  $\omega$ and we are left with the following integration over positive imaginary frequencies $\omega=\rmi \xi$:
\begin{equation} \label{EE}
\langle\mathbf{\hat{E}}(\mathbf{r},t)\otimes\mathbf{\hat{E}}(\mathbf{r},t)\rangle=-\frac{\hbar\mu_0}{\pi}\int_{0}^\infty\mathbf{G}^S(\mathbf{r},\mathbf{r},\rmi \xi)\,\xi^2 \rmd\xi.
\end{equation}
The equal-time correlation function for the magnetic field operator is obtained in a similar manner using (\ref{EBdef}) and (\ref{As}), with the result
\begin{equation} \label{BB}
\langle\mathbf{\hat{B}}(\mathbf{r},t)\otimes\mathbf{\hat{B}}(\mathbf{r},t)\rangle=\frac{\hbar\mu_0}{\pi}\lim_{\mathbf{r}\rightarrow\mathbf{r'}}\int_{0}^\infty\nabla\times\mathbf{G}^S(\mathbf{r},\mathbf{r'},\rmi \xi)\times\stackrel{\leftarrow}{\nabla'}\, \rmd\xi,
\end{equation}
where $\times\stackrel{\leftarrow}{\nabla'}$ denotes a curl on the second index of the Green tensor, so that for a vector $\mathbf{V}(\mathbf{r'})$ we have $\mathbf{V}\times\stackrel{\leftarrow}{\nabla'}=\nabla'\times\mathbf{V}$.
The fact that the integrations in (\ref{EE}) and (\ref{BB}) are over imaginary frequncies ($\omega=\rmi \xi$) ensures that the integrals are well behaved~\cite{lif55,LL,dzy61}.  The stress tensor (\ref{stress}) follows immediately from (\ref{EE}) and (\ref{BB}).

The expressions (\ref{EE}) and ({\ref{BB}) are familiar from Lifshitz theory~\cite{lif55,dzy61,LL}, except that here the symmetrized Green tensor $\mathbf{G}^S$ appears in these correlation functions rather than $\mathbf{G}$. In the static case of non-moving plates there is in fact no difference since the Green tensor is diagonal when $\mathbf{r}=\mathbf{r'}$ (in the Cartesian coordinates of Figure~\ref{fig1}). For the problem considered here, however, the emergence of the symmetrized Green tensor in  (\ref{EE}) and ({\ref{BB}) ensures that the electromagnetic stress tensor is symmetric. We briefly return to this point in Section~\ref{Stress}.

Because of the interaction of the modes with the plates, which are in general dissipative, one may object that they cannot therefore be normalized as in (\ref{norms}). This technical issue of quantum electrodynamics in absorptive media can be dealt with using Schwinger's source theory and its modern version~\cite{abs}; here we used the modes as a tool to obtain (\ref{EE}) and ({\ref{BB}), which remain valid in absorptive media~\cite{abs}. The issue of dissipation has also been much discussed in the Casimir literature (see~\cite{moc06} and references therein) with the result that expressions for the Casimir force in terms of reflection and transmission coefficients (such as are derived here) are valid for dissipative media. This is intuitively reasonable since the quantum vacuum persists even in absorptive materials.

\section{The Green tensor}  \label{Green}
With retarded boundary conditions, the solution of the mono\-chromatic equation (\ref{green}) has a simple physical meaning: an oscillating dipole at the point $\mathbf{r'}$ emits electromagnetic waves of frequency $\omega$ and $\mathbf{G}(\mathbf{r},\mathbf{r'},\omega,\beta)$ is the resulting vector potential at the point $\mathbf{r}$. The second index in $G_{ij}$ represents the orientation of the dipole at $\mathbf{r'}$, while the first index represents the components of the vector potential at $\mathbf{r}$. Using this physical consideration it is clear from Fig.~\ref{fig1} that the solution will be a linear superposition of waves that have reflected off the plates, with the number of reflections ranging from zero to infinity. To write down the solution we exploit the homogeneity of the problem in the $y$- and $z$- directions to Fourier transform the Green tensor as follows:
\begin{equation} \label{four}
\mathbf{\widetilde{G}}(x,x',u,v,\rmi \xi) 
=\int_{-\infty}^\infty \rmd y\int_{-\infty}^\infty \rmd z\,\mathbf{G}(\mathbf{r},\mathbf{r'},\rmi \xi)\,\rme^{-\rmi u(y-y')-\rmi v(z-z')}
\end{equation}
so that we decompose the waves emitted by the dipole into plane waves. In the absence of the plates the solution is the bare Green tensor~\cite{LL}
\begin{eqnarray}
\mathbf{\widetilde{G}}_b(x,x',u,v,\rmi \xi)=
\left\{\begin{array}{l} \rme^{-w(x-x')}\boldsymbol{\mathcal{G}_+}, \quad x>x'  \\[5pt] \rme^{w(x-x')}\boldsymbol{\mathcal{G}_-}, \quad x<x' \end{array}\right.,  \label{bare} \\[6pt]
\boldsymbol{\mathcal{G}_\pm}=-\frac{1}{2w\kappa^2}\left[\left(\begin{array}{c} \pm w \\ \rmi u \\ \rmi v \end{array}\right)\!\!\otimes\!\!\left(\begin{array}{c} \pm w \\ \rmi u \\ \rmi v \end{array}\right)-\kappa^2\mathds{1}\right],\\[6pt]
 \kappa=\frac{\xi}{c}, \qquad w=\sqrt{u^2+v^2+\kappa^2}. \label{w}
\end{eqnarray}
The two possibilities in (\ref{bare}) are plane waves propagating to the right (first line) or to the left (second line), with wave vectors $\mathbf{k}=(\pm \rmi w,u,v)$.
The imaginary $x$-component of the wave vectors is a consequence of the imaginary frequency, and 
in (\ref{w}) we simply have the relation $\omega=ck$. In physical terms the vacuum solution (\ref{bare}) is trivial: it is the only way the dipole can propagate plane waves from $x'$ to $x$. In the presence of the plates both plane waves in (\ref{bare}) will reflect off the plates and reverse direction, so the left-moving plane wave can propagate from $x'$ to $x$ even if $x>x'$, with similar considerations applying to the right-moving plane wave. Consequently, both the right- and left-moving waves will appear in the solution regardless of whether $x$ is greater or less than $x'$, in contrast to the vacuum solution (\ref{bare}). Let $\mathbf{R}_2$ be the reflection operator (matrix) that transforms a right-moving plane wave at plate~2 into the resulting reflected left-moving plane wave, and let $\mathbf{R}_1$ be the reflection operator that transforms a left-moving plane wave at plate~1 into the reflected right-moving plane wave. We can now write down the solution (functional dependences are suppressed):
\begin{eqnarray} 
\fl  \mathbf{\widetilde{G}}=\mathbf{\widetilde{G}}_b-\rme^{-w(x-x')}\boldsymbol{\mathcal{G}_+}-\rme^{w(x-x')}\boldsymbol{\mathcal{G}_-}  \nonumber \\
 +\left(\mathds{1}-\rme^{-2wa}\mathbf{R}_1\mathbf{R}_2\right)^{-1} 
   \left(\rme^{-w(x-x')}\boldsymbol{\mathcal{G}_+}+\rme^{-w(x+x')}\mathbf{R}_1\boldsymbol{\mathcal{G}_-}\right)\nonumber  \\
 +\left(\mathds{1}-\rme^{-2wa}\mathbf{R}_2\mathbf{R}_1\right)^{-1} 
   \left(\rme^{w(x-x')}\boldsymbol{\mathcal{G}_-}+\rme^{w(x+x'-2a)}\mathbf{R}_2\boldsymbol{\mathcal{G}_+}\right). \label{soln}
\end{eqnarray}
The first line in (\ref{soln}) subtracts the left- or right-moving plane wave in (\ref{bare}), depending on whether $x$ is greater or less than $x'$. This subtraction is necessary because the direct propagation, without reflections, of both the right- and left-moving plane waves from $x'$ to $x$ is contained in the remaining terms in (\ref{soln}); but only one of these propagations is possible, depending on whether $x$ is greater or less than $x'$, and the first line in (\ref{soln}) automatically subtracts the irrelevant one. The inverse matrices in (\ref{soln}) are geometric series representing every possible number of double reflections off both plates, the exponentials providing the propagation distance $2a$ for each double reflection. The initial right- and left-moving plane waves that leave $x'$ reach $x$ after both an even and odd number of reflections; this explains the terms multiplying the inverse matrices in (\ref{soln}). Each term in (\ref{soln}), after the series expansion of the inverse matrices, has  an overall exponential factor that accounts for the propagation distance involved, with $\rme^{-ws}$, $s>0$, representing a propagation distance $s$ to the right for the initial right-moving plane wave, but to the left for the initial left-moving plane wave. Note that we have not omitted any part of the derivation of (\ref{soln}); it was written down using the reasoning described above.

It remains to deduce the reflection operators $\mathbf{R}_1$ and $\mathbf{R}_2$. Reflection of left-moving plane waves at plate~1 can be calculated by decomposing the plane wave into two components, one with polarization in the plane of incidence (TM) and the other with polarization perpendicular to this plane (TE); these components must then be multiplied by the well-known reflection coefficients for these polarizations~\cite{jac}. As well as the overall change in amplitude of the two polarization components, the $x$-component of $\mathbf{A}$ (proportional to $\mathbf{E}$) changes sign on reflection, to remain perpendicular to $\mathbf{k}$. The reflection operator $\mathbf{R}_1$ is therefore
\begin{eqnarray}
\mathbf{R}_1=r_{E1}\mathbf{P}_{E1}+r_{B1}\mathbf{R}_x\mathbf{P}_{B1}, \label{R1}   \\[5pt] 
\mathbf{R}_x=\left(\begin{array}{ccc} -1 & 0 & 0 \\ 0 & 1 & 0 \\ 0 & 0 & 1 \end{array}\right),  \label{Rx}  \\[5pt]  
r_{E1}=\frac{\mu_1(\rmi c\kappa)w-w_1}{\mu_1(\rmi c\kappa)w+w_1}, \quad r_{B1}=-\frac{\varepsilon_1(\rmi c\kappa)w-w_1}{\varepsilon_1(\rmi c\kappa)w+w_1}, \label{refs1}   \\ 
w_1=\sqrt{u^2+v^2+\varepsilon_1(\rmi c\kappa)\mu_1(\rmi c\kappa)\kappa^2}.
\end{eqnarray}
In (\ref{R1}), $\mathbf{P}_{B1}$ ($\mathbf{P}_{E1}$) projects $\mathbf{E}$ to its component in (perpendicular to) the plane of incidence, while $\mathbf{R}_x$ flips the $x$-component of $\mathbf{E}$ (the $E$-polarization has no $x$-component). Eqns.~(\ref{refs1}) are the reflection coefficients~\cite{jac}. The (left-moving) wave vectors in the vacuum ($\mathbf{k}$) and the plate ($\mathbf{k}_1$) are
\begin{equation} \label{lks}
\mathbf{k}=(-\rmi w,u,v), \qquad \mathbf{k}_1=(-\rmi w_1,u,v).
\end{equation}
The projection operators are given by
\begin{eqnarray}
\mathbf{P}_{E1}=\mathbf{n}_{E1}\otimes\mathbf{n}_{E1}, \qquad \mathbf{P}_{B1}=\mathbf{n}_{B1}\otimes\mathbf{n}_{B1}, \\[5pt]
\mathbf{n}_{E1}=\frac{1}{\sqrt{u^2+v^2}}\left(\begin{array}{c} 0 \\ -v \\ u \end{array}\right),  \qquad
\mathbf{n}_{B1}=\frac{1}{\kappa\sqrt{u^2+v^2}}\left(\begin{array}{c} \rmi (u^2+v^2) \\ -uw \\ -vw \end{array}\right), \label{endR1}
\end{eqnarray}
where $\mathbf{n}_{E1}$ and $\mathbf{n}_{B1}$ are unit vectors in the two polarization directions, for a left-moving plane wave.

If $\beta=0$, the reflection operator $\mathbf{R}_2$ for right-moving plane waves at plate~2 is similar to (\ref{R1})--(\ref{endR1}). As the waves are now right-moving, instead of (\ref{lks}) we have the wave vectors
\[
\mathbf{k}=(\rmi w,u,v), \qquad \mathbf{k}_2=(\rmi w_2,u,v).
\]
We therefore simply change the material quantities in (\ref{R1})--(\ref{endR1}) to those of plate~2 and flip the sign of the $w$'s (this affects the polarization direction (\ref{endR1})). When $\beta\neq0$ this result for plate~2 holds in an inertial frame co-moving with the plate. To obtain  $\mathbf{R}_2$ in the frame of Fig.~\ref{fig1}, we must Lorentz boost in the negative $y$-direction. Denoting by primes any quantities in the co-moving frame that differ from their values in the frame of Fig.~\ref{fig1}, we have~\cite{jac}
\begin{eqnarray}
\kappa'=\gamma(\kappa+\rmi \beta u), \quad u'=\gamma(u-\rmi \beta\kappa),  \quad \gamma=(1-\beta^2)^{-\frac{1}{2}}, \label{ktrans} \\
E'_x=\gamma(E_x+\beta cB_z), \qquad E'_z=\gamma(E_z-\beta cB_x), \label{lor1} \\
B'_x=\gamma(B_x-\beta E_z/c), \qquad B'_z=\gamma(B_z+\beta E_x/c). \label{lor2}
\end{eqnarray}
This mixing of the $\mathbf{E}$ and $\mathbf{B}$ fields gives the required transformation of the polarization directions. In this manner we find $\mathbf{R}_2$ in the frame of Fig.~\ref{fig1}:
\begin{eqnarray}
\mathbf{R}_2=r_{E2}\mathbf{R}_x\mathbf{P}_{E2}+r_{B2}\mathbf{R}_x\mathbf{P}_{B2}, \label{R2}   \\[5pt]  
r_{E2}=\frac{\mu_2(\rmi c\kappa')w-w_2}{\mu_2(\rmi c\kappa')w+w_2}, \quad r_{B2}=-\frac{\varepsilon_2(\rmi c\kappa')w-w_2}{\varepsilon_2(\rmi c\kappa')w+w_2}, \label{refs2}  \\ 
w_2=\sqrt{u'^2+v^2+\varepsilon_2(\rmi c\kappa')\mu_2(\rmi c\kappa')\kappa'^2},   \\
\mathbf{P}_{E2}=\mathbf{n}_{E2}\otimes\mathbf{n}_{E2}, \qquad \mathbf{P}_{B2}=\mathbf{n}_{B2}\otimes\mathbf{n}_{B2},  \\
\mathbf{n}_{E2}=\frac{1}{\kappa\sqrt{u^2+v^2-2\rmi \beta\kappa u-\beta^2(\kappa^2+v^2)}} 
\left(\begin{array}{c} \beta vw \\ -v(\kappa+\rmi \beta u) \\ \kappa u-\rmi \beta(\kappa^2+v^2) \end{array}\right), 
 \\[5pt]
\mathbf{n}_{B2}=\frac{1}{\kappa\sqrt{u^2+v^2-2\rmi\beta\kappa u-\beta^2(\kappa^2+v^2)}} 
\left(\begin{array}{c} \rmi (u^2+v^2)+\beta\kappa u \\ w(u-\rmi \beta\kappa) \\ vw \end{array}\right). 
\label{endR2}
\end{eqnarray}
Note that the reflection coefficients (\ref{refs2}) are evaluated using quantities in the frame co-moving with plate~2.

We have verified the solution (\ref{soln})--(\ref{endR2}) by comparing it with a brute-force numerical solution of the boundary-value problem for the Green tensor. A novel feature of our approach is that the exact analytical solution is obtained purely by physical reasoning, without directly solving the differential equation (\ref{green}).

\section{The stress tensor}  \label{Stress}
We can now compute the stress tensor. It is first necessary to drop the bare Green tensor (\ref{bare}) in (\ref{soln}), as this gives the diverging zero-point stress in the absence of the plates~\cite{LL}.  The remaining part of the symmetrized Green tensor (\ref{GS}) gives the expectation values (\ref{EE})-- (\ref{BB}), which determine the stress tensor (\ref{stress}). Some labour is required, as the inverse matrices in (\ref{soln}) must be evaluated and simplified; the details are given in~\ref{A1}. The stress tensor is diagonal, so there is no lateral force on the plates. The component $\sigma_{xx}$ of the stress is constant between the plates and is equal to the perpendicular force per unit area $F$ on the plates. A positive sign for $F$ means the plates are attracted to each other, whereas a negative sign means they repell each other. We introduce the quantities
\begin{eqnarray*}
A_{EE}=r_{E1}^{-1}r_{E2}^{-1}\rme^{2aw}-1, \qquad A_{BB}=r_{B1}^{-1}r_{B2}^{-1}\rme^{2aw}-1, \\
A_{EB}=r_{E1}^{-1}r_{B2}^{-1}\rme^{2aw}-1, \qquad A_{BE}=r_{B1}^{-1}r_{E2}^{-1}\rme^{2aw}-1,
\end{eqnarray*}
and write the force per unit area:
\begin{eqnarray} 
\fl  F=\frac{\hbar c}{4\pi^3}\int_0^\infty \rmd\kappa\int_{-\infty}^\infty \rmd u\int_{-\infty}^\infty \rmd v\,w  \nonumber \\
\times \left[\frac{(A_{EE}+A_{BB})(u^2+v^2-\rmi \kappa u\beta)^2-(A_{EB}+A_{BE})w^2v^2\beta^2}{A_{EE}A_{BB}(u^2+v^2-\rmi \kappa u\beta)^2-A_{EB}A_{BE}w^2v^2\beta^2}\right].   \label{force}
\end{eqnarray}
Note that the numerator in (\ref{force}) becomes the denominator if the sums of pairs of $A$'s are replaced by products. Despite the presence of imaginary terms in the integrand, the expression (\ref{force}) is a real number. This is easily seen by considering a Taylor expansion in $u$ of the integrand. Since the permittivities and permeabilities are real on the positive imaginary frequency axis~\cite{LLcm}, this expansion consists of real terms even in $u$ and imaginary terms odd in $u$; the latter vanish after the integration with respect to $u$. When $\beta=0$, Eqn.~(\ref{force}) reduces to Lifshitz's result~\cite{lif55,LL}
\begin{equation} \label{Lforce}
F(\beta=0)=\frac{\hbar c}{4\pi^3}\int_0^\infty \rmd\kappa\int_{-\infty}^\infty \rmd u\int_{-\infty}^\infty \rmd v\,w \left(A_{EE}^{-1}+A_{BB}^{-1}\right).
\end{equation}

The perpendicular force clearly cannot depend on the sign of $\beta$, so the lowest order correction to (\ref{Lforce}) when $\beta\neq0$ is proportional to $\beta^2$ (a lateral force could of course be proportional to $\beta$ to lowest order). The expression for this $O(\beta^2)$ correction to (\ref{Lforce}) depends on the dispersion in Plate~2, as  $\varepsilon_2$ and $\mu_2$ in the reflection coefficients  (\ref{refs2}) are evaluated at the co-moving (imaginary) frequency given in (\ref{ktrans}); one must therefore also expand $r_{E2}$ and $r_{B2}$ to  $O(\beta^2)$ to obtain the required correction. If we assume there is no dispersion then the perpendicular force (\ref{force}) expanded to  $O(\beta^2)$  is
\begin{eqnarray} 
\fl  F(O(\beta^2))=\frac{\hbar c}{4\pi^3}\int_0^\infty \rmd\kappa\int_{-\infty}^\infty \rmd u\int_{-\infty}^\infty \rmd v\,w \left[A_{EE}^{-1}+A_{BB}^{-1}\right. \nonumber \\
\left.+\beta^2\frac{w^2v^2e^{2aw}(r_{E1}-r_{B1})(r_{E2}-r_{B2})\left(e^{4aw}-r_{E1}r_{E2}r_{B1}r_{B2}\right)}{(u^2+v^2)^2\,(r_{E1}r_{E2}r_{B1}r_{B2}A_{EE}A_{BB})^{2}}\right].  \label{Lforcecor}
\end{eqnarray}

The Casimir force (\ref{Lforce}) between non-moving plates is attractive ($F>0$) for dielectrics but can be repulsive ($F<0$) if there is a magnetic response. If each plate has either an infinite permittivity or an infinity permeability, (\ref{force}) and (\ref{Lforce}) give the same result so the motion has no effect on the force. For general dispersive dielectrics we find that the motion serves to add an attractive component to the force compared to the non-moving case, whereas with a magnetic response this component can be attractive or repulsive.

Presentations of the static problem~\cite{LL} ($\beta=0$) do not usually address the issue of the electromagnetic stress tensor inside the plates. It turns out that the relevant component of the stress tensor (the $xx$-component) is zero in the plates in the static case so the correct Casimir force can be found by restricting the analysis to the stress between the plates. (The $yy$- and $zz$-components of the stress in the material are not zero, however, giving a non-vanishing Casimir energy in the plates.) It is important to check whether this situation is maintained in the moving case $\beta\neq 0$. To find the stress tensor in Plate~1 we require the Green tensor for the wave equation in the material. The bare Green tensor is a simple generalization of (\ref{bare}) and the physical reasoning used above to write down the Green tensor between the plates can readily be applied to find the Green tensor in the material by moving $\mathbf{r}$ and $\mathbf{r'}$ in Figure~1 into Plate~1. The details are presented in~\ref{A2} and the results show that the electromagnetic stress tensor in Plate~1 is diagonal with zero $xx$-component, as in the static case; hence the complete Casimir force on the plate is indeed the perpendicular force (\ref{force}). By Lorentz invariance the same applies to Plate~2.

As mentioned in the Introduction the previous attempts~\cite{vol99} at an exact solution of this problem failed to calculate the Green tensor correctly (moreover, the important issue of the stress tensor in the plates was not addressed). The expressions for the Green tensor and any non-vanishing vacuum forces cannot be written as a sum of contributions from two orthogonal polarizations as in the static case (\ref{Lforce}). In~\cite{vol99} it was mistakenly concluded that this feature of the static result would be preserved in the moving case (this error was also made in~\cite{pen97}). As we have demonstrated, the relevant polarization decomposition for a reflecting wave is different at each plate, even to first order in $\beta$, due to the Lorentz transformation (\ref{lor1})--(\ref{lor2}); one can only use the same polarization decomposition at each plate if one ignores the motion. There are therefore four, rather than two,  polarization directions that have to be considered for each plane wave. The lack of simple orthogonality relations between these four directions is the reason why the correct solution (\ref{force}) for $\beta\neq 0$ looks nothing like (\ref{Lforce}). Nevertheless, given the complexity of the Green tensor from which it is calculated (see~\ref{A1}), the expression (\ref{force}) has a certain elegance of its own.

\section{Conclusion}
Shear motion at a constant speed of one infinite plate parallel to another modifies the Casimir force between them compared to the non-moving case, but it does not induce a lateral force on the plates. It has been shown~\cite{dzy61} that the Casimir-Polder force between a polarizable particle and an infinite plate, at rest relative to each other, can be obtained from the Lifshitz formula (\ref{Lforce}) for the force between two infinite plates by a limiting procedure wherein one of the plates is allowed to become rarified. Alternatively~\cite{sch78}, the Green tensor outside a single plate can be obtained as a limit of the two-plate Green tensor and used to calculate the force on a polarizable particle. It is clear from both these methods that our result implies the absence of ``quantum friction" on a particle moving a constant speed parallel to an infinite plate. 

Very different is the case of two finite bodies moving past each other with constant velocities: this is a complicated dynamical problem, even if the effect of the vaccum forces on the velocities of the bodies is neglected. But it is questionable if some feature of the interaction of the bodies could be usefully singled out as ``friction", especially if the separation of the bodies is not small compared to their sizes. In particular, depending on the state of motion, a net attractive force between dielectric bodies could act to speed them up rather than slow them down.

\ack
This research is supported by the Royal Society of Edinburgh, the Scottish Government, the Leverhulme Trust, the Max Planck Society, the National University of Singapore and a Royal Society Wolfson Research Merit Award.

\appendix 
\section{Green tensor between the plates: expanded form}   \label{A1}
The Green tensor given by (\ref{soln})--(\ref{endR2}) contains the inverse matrices
\begin{equation} \label{mat1s}
\left(\mathds{1}-\rme^{-2wa}\mathbf{R}_1\mathbf{R}_2\right)^{-1} \quad \mathrm{and} \quad
 \left(\mathds{1}-\rme^{-2wa}\mathbf{R}_2\mathbf{R}_1\right)^{-1},
\end{equation}
which must be expanded before computing the stress tensor. These matrices clearly have a geometric-series expansion and this, together with the form of the reflection operators $\mathbf{R}_1$ and $\mathbf{R}_2$ (see (\ref{R1}) and (\ref{R2})) shows that they can be written in a basis constructed from the vectors $\mathbf{n}_{E1}$, $\mathbf{n}_{B1}$, $\mathbf{n}_{E2}$, and $\mathbf{n}_{B2}$. A little thought leads one to conclude that
\begin{eqnarray}
  \left(\mathds{1}-\rme^{-2wa}\mathbf{R}_1\mathbf{R}_2\right)^{-1}=\mathds{1}+\mathbf{M}_2, \label{inv1s} \\
\fl  \mathbf{M}_2=c_{2EE}\,\mathbf{n}_{E2}\otimes\mathbf{n}_{E2}+c_{2BB}\,\mathbf{n}_{B2}\otimes\mathbf{n}_{B2}+c_{2EB}\,\mathbf{n}_{E2}\otimes\mathbf{n}_{B2}+c_{2BE}\,\mathbf{n}_{B2}\otimes\mathbf{n}_{E2}, \label{inv2s} \\
\left(\mathds{1}-\rme^{-2wa}\mathbf{R}_2\mathbf{R}_1\right)^{-1}=\mathds{1}+\mathbf{M}_1, \label{inv3s} \\
\fl  \mathbf{M}_1=c_{1EE}\,\mathbf{n}_{E1}\otimes\mathbf{n}_{E1}+c_{1BB}\,\mathbf{n}_{B1}\otimes\mathbf{n}_{B1}+c_{1EB}\,\mathbf{n}_{E1}\otimes\mathbf{n}_{B1}+c_{1BE}\,\mathbf{n}_{B1}\otimes\mathbf{n}_{E1}, \label{inv4s}
\end{eqnarray}
where the various $c$'s are unknown coefficients. At this stage (\ref{inv1s})--(\ref{inv4s}) represent a conjecture; it will be verified at the end of the procedure that they are true. The unknown $c$-coefficients in (\ref{inv2s}) and (\ref{inv4s}) can be found by taking successive dot products of (\ref{mat1s}) with polarization vectors; to give one example, $c_{2EB}$ is given by
\begin{equation} \label{dots}
c_{2EB}=\left[\left(\mathds{1}-\rme^{-2wa}\mathbf{R}_1\mathbf{R}_2\right)^{-1}-\mathds{1}\right]_{ij}(\mathbf{n}_{B2})_j(\mathbf{n}_{E2})_i.
\end{equation}
We used Mathematica to calculate all of the required dot products of the form (\ref{dots}) and thereby found all of the unknown $c$-coefficients. Defining $\alpha$, $\lambda$ and $\nu$ by
\begin{eqnarray}
\fl  \alpha=\left[\rme^{4aw}+r_{E1}r_{E2}r_{B1}r_{B2}-\rme^{2aw}(r_{E1}r_{E2}\lambda^2+r_{B1}r_{B2}\lambda^2+r_{E2}r_{B1}\nu^2+r_{E1}r_{B2}\nu^2)\right]^{-1}, \\
\fl  \lambda=\mathbf{n}_{E1}\cdot\mathbf{n}_{E2}=\frac{u^2+v^2-\rmi \kappa u\beta}{\sqrt{u^2+v^2}\sqrt{u^2+v^2-2\rmi \kappa u\beta-(\kappa^2+v^2)\beta^2}}, \qquad
\nu^2=1-\lambda^2,
\end{eqnarray}
we write the results for the $c$-coefficients:
\begin{eqnarray}
c_{2EE}=\alpha r_{E2}\left[-r_{E1}r_{B1}r_{B2}+\rme^{2aw}(r_{E1}\lambda^2+r_{B1}\nu^2)\right],  \\
 c_{2BB}=\alpha r_{B2}\left[-r_{E1}r_{E2}r_{B1}+\rme^{2aw}(r_{B1}\lambda^2+r_{E1}\nu^2)\right], \\
 c_{1EE}=\alpha r_{E1}\left[-r_{E2}r_{B1}r_{B2}+\rme^{2aw}(r_{E2}\lambda^2+r_{B2}\nu^2)\right],  \\
 c_{1BB}=\alpha r_{B1}\left[-r_{E1}r_{E2}r_{B2}+\rme^{2aw}(r_{B2}\lambda^2+r_{E2}\nu^2)\right],  \\
\fl   c_{2BE}=\alpha \rme^{2aw}r_{E2}(r_{E1}-r_{B1})\lambda\nu, \qquad
c_{2EB}=\alpha \rme^{2aw}r_{B2}(r_{E1}-r_{B1})\lambda\nu, \\
\fl   c_{1BE}=\alpha \rme^{2aw}r_{E1}(r_{E2}-r_{B2})\lambda\nu, \qquad
c_{1EB}=\alpha \rme^{2aw}r_{B1}(r_{E2}-r_{B2})\lambda\nu.
\end{eqnarray}
The expanded form (\ref{inv1s})--(\ref{inv4s}) of the inverse matrices (\ref{mat1s}) is now determined and the final step in the procedure was to verify using Mathematica that (\ref{inv1s})--(\ref{inv4s}) are indeed correct.

From (\ref{soln}) we see that the Green tensor also contains
\begin{equation} \label{mat2s}
\left(\mathds{1}-\rme^{-2wa}\mathbf{R}_1\mathbf{R}_2\right)^{-1}\mathbf{R}_1 \quad \mathrm{and} \quad
 \left(\mathds{1}-\rme^{-2wa}\mathbf{R}_2\mathbf{R}_1\right)^{-1}\mathbf{R}_2.
\end{equation}
Using the above results for (\ref{mat1s}), these can be written in the form
\begin{eqnarray}
\left(\mathds{1}-\rme^{-2wa}\mathbf{R}_1\mathbf{R}_2\right)^{-1}\mathbf{R}_1=\mathbf{R}_x\mathbf{N}_1, \\
\fl   \mathbf{N}_1=d_{1EE}\,\mathbf{n}_{E1}\otimes\mathbf{n}_{E1}+d_{1BB}\,\mathbf{n}_{B1}\otimes\mathbf{n}_{B1}+d_{1EB}\,\mathbf{n}_{E1}\otimes\mathbf{n}_{B1}+d_{1BE}\,\mathbf{n}_{B1}\otimes\mathbf{n}_{E1}, \\
\left(\mathds{1}-\rme^{-2wa}\mathbf{R}_2\mathbf{R}_1\right)^{-1}\mathbf{R}_2=\mathbf{R}_x\mathbf{N}_2, \\
\fl   \mathbf{N}_2=d_{2EE}\,\mathbf{n}_{E2}\otimes\mathbf{n}_{E2}+d_{2BB}\,\mathbf{n}_{B2}\otimes\mathbf{n}_{B2}+d_{2EB}\,\mathbf{n}_{E2}\otimes\mathbf{n}_{B2}+d_{2BE}\,\mathbf{n}_{B2}\otimes\mathbf{n}_{E2},
\end{eqnarray}
where $\mathbf{R}_x$ is defined by (\ref{Rx}) and the $d$-coefficients are
\begin{eqnarray}
d_{1EE}=\alpha \rme^{2aw}r_{E1}\left[\rme^{2aw}-r_{B1}(r_{B2}\lambda^2+r_{E2}\nu^2)\right], \\
d_{1BB}=\alpha \rme^{2aw}r_{B1}\left[\rme^{2aw}-r_{E1}(r_{E2}\lambda^2+r_{B2}\nu^2)\right], \\
d_{1BE}=d_{1EB}=\alpha \rme^{2aw}r_{E1}r_{B1}(r_{E2}-r_{B2})\lambda\nu, \\
d_{2EE}=\alpha \rme^{2aw}r_{E2}\left[\rme^{2aw}-r_{B2}(r_{B1}\lambda^2+r_{E1}\nu^2)\right], \\
d_{2BB}=\alpha \rme^{2aw}r_{B2}\left[\rme^{2aw}-r_{E2}(r_{E1}\lambda^2+r_{B1}\nu^2)\right], \\
d_{2BE}=d_{2EB}=\alpha \rme^{2aw}r_{E2}r_{B2}(r_{E1}-r_{B1})\lambda\nu.
\end{eqnarray}
Finally, we can write the Green tensor (\ref{soln}) in expanded form:
\begin{eqnarray} 
\fl   \mathbf{\widetilde{G}}=\mathbf{\widetilde{G}}_b+\left[\rme^{-w(x-x')}\mathbf{M}_2+\rme^{w(x+x'-2a)}\mathbf{R_x}\mathbf{N}_2\right]\boldsymbol{\mathcal{G}_+}   \nonumber  \\
+\left[\rme^{w(x-x')}\mathbf{M}_1+\rme^{-w(x+x')}\mathbf{R_x}\mathbf{N}_1\right]\boldsymbol{\mathcal{G}_-}.   \label{Gsims}
\end{eqnarray}
We also verified directly using Mathematica that (\ref{Gsims}) is equal to (\ref{soln}). In the form (\ref{Gsims}) the Green tensor was used to calculate the electromagnetic stress tensor as described in the main text.

\section{The electromagnetic stress tensor in Plate~1}  \label{A2}
In this Appendix we discuss the Green tensor for the vector potential inside Plate~1 and the resulting electromagnetic stress tensor. In Plate~1 the Green tensor satisfies (we use a subscript $m$ for the Green tensor in the material)
\begin{eqnarray}
\mathbf{G}_m(\mathbf{r},\mathbf{r'},\omega)=\int^\infty_{0}\rmd t \,\mathbf{G}_m(\mathbf{r},t,\mathbf{r'},0)\,\rme^{\rmi \omega t},   \label{freqGms} \\[5pt]
\left(\nabla\times\frac{1}{\mu_1(\omega)}\nabla\times-\frac{\omega^2}{c^2}\varepsilon_1(\omega)\right)\mathbf{G}_m(\mathbf{r},\mathbf{r'},\omega)=\mathds{1}\delta(\mathbf{r}-\mathbf{r'}), \label{greenms}
\end{eqnarray}
The bare Green tensor in the plate is that for a material filling all space, with no material boundaries. In terms of the Fourier transform
\begin{equation} \label{fourms}
\mathbf{\widetilde{G}}_m(x,x',u,v,\rmi \xi)
=\int_{-\infty}^\infty \rmd y\int_{-\infty}^\infty \rmd z\,\mathbf{G}_m(\mathbf{r},\mathbf{r'},\rmi \xi)\,\rme^{-\rmi u(y-y')-\rmi v(z-z')}
\end{equation}
the bare Green tensor is~\cite{dzy61}
\begin{eqnarray}
\mathbf{\widetilde{G}}_{mb}(x,x',u,v,\rmi \xi)=
\left\{\begin{array}{l} \rme^{-w_1(x-x')}\boldsymbol{\mathcal{G}_{m+}}, \quad x>x'  \\[5pt] \rme^{w_1(x-x')}\boldsymbol{\mathcal{G}_{m-}}, \quad x<x' \end{array}\right., \\
\boldsymbol{\mathcal{G}_{m\pm}}=-\frac{1}{2\varepsilon_1 w_1\kappa^2}\left[\left(\begin{array}{c} \pm w_1 \\ \rmi u \\ \rmi v \end{array}\right)\!\!\otimes\!\!\left(\begin{array}{c} \pm w_1 \\ \rmi u \\ \rmi v \end{array}\right)-\varepsilon_1\mu_1\kappa^2\mathds{1}\right], \label{Gcalms} \\[6pt]
 \kappa=\frac{\xi}{c}, \qquad w_1=\sqrt{u^2+v^2+\varepsilon_1\mu_1\kappa^2}. \label{ws}
\end{eqnarray}
The two possibilities in (\ref{Gcalms}) are plane waves propagating to the right (first line) or to the left (second line), with wave vectors $\mathbf{k}_1=(\pm iw_1,u,v)$.

\begin{figure}
\begin{center}
\includegraphics[width=15.0pc]{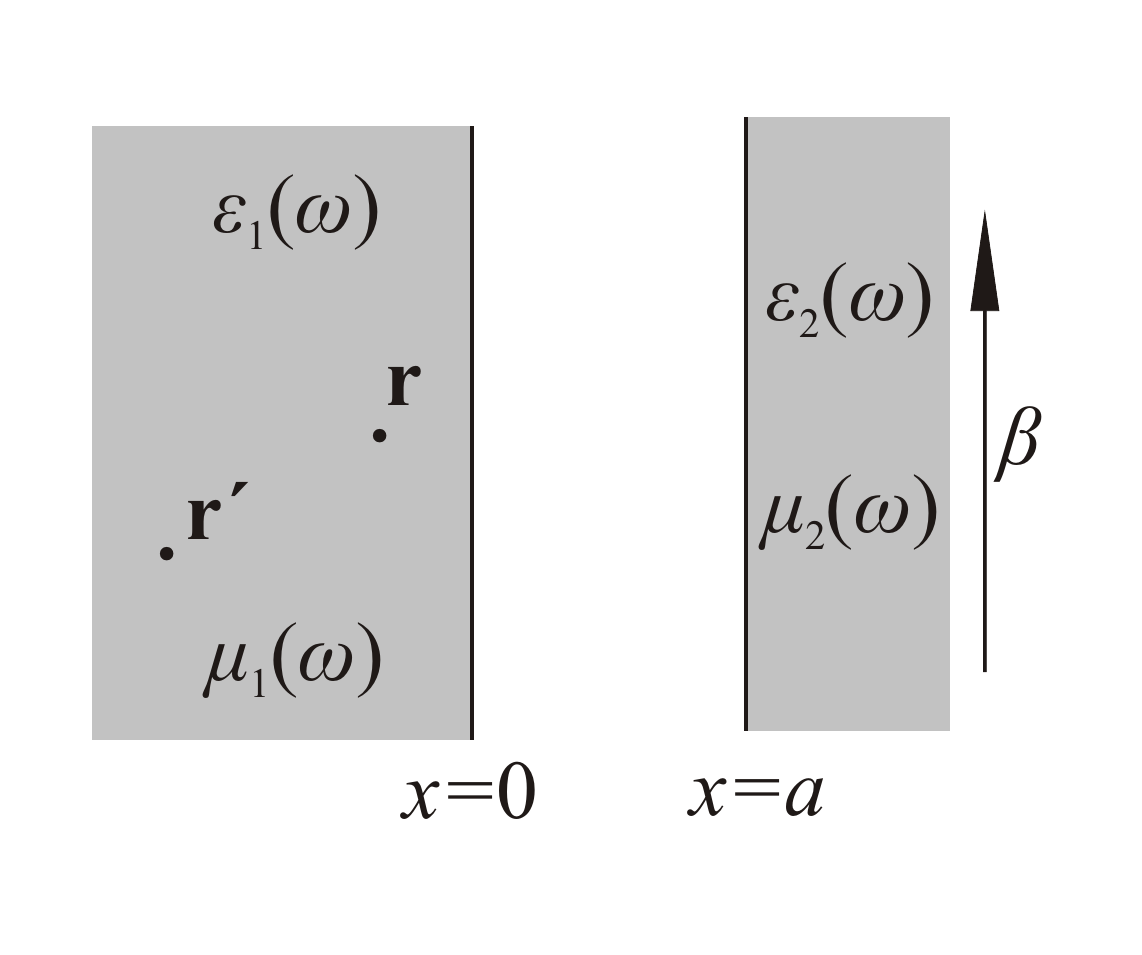}
\vspace*{-7mm}
\caption{Calculation of the Green tensor in Plate~1. \label{fig3}}
\end{center}
\end{figure}
The Green tensor $\mathbf{G}_m(\mathbf{r},t,\mathbf{r'},t')$ represents the vector potential at  a point $\mathbf{r}$ in Plate~1 resulting from a radiating dipole at another point $\mathbf{r'}$ in the plate (see Figure~\ref{fig3}). Because of the Fourier decomposition (\ref{fourms}) we need only consider plane waves. As in Section~\ref{Green} we can simply write down the solution for the Green tensor. Let $\mathbf{R}_m$ be the reflection operator for a right-moving plane wave in the plate at the $x=0$ boundary and let $\mathbf{T}_m$ be the transmission operator for that plane wave at the  $x=0$ boundary. Let $\mathbf{T}_1$ be the transmission operator for a left-moving plane wave in the vacuum between the plates at the $x=0$ boundary. The bare Green tensor (\ref{Gcalms}) represents direct propagation from $x'$ to $x$. A plane wave can also propagate from $x'$ to $x$ by reflecting off the boundary at $x=0$, or by being transmitted through the $x=0$ boundary, reflecting once off Plate~2, then reflecting any even number (including zero) of times off both plates before finally being transmitted at the $x=0$ boundary. The Green tensor is therefore
\begin{eqnarray} \label{Gm1s}
\fl   \mathbf{\widetilde{G}}_m=\mathbf{\widetilde{G}}_{mb}+\rme^{w_1(x'+x)}\mathbf{R}_m\boldsymbol{\mathcal{G}_{m+}}  \\
+\rme^{w_1(x'+x)}\mathbf{T}_1\left(\mathds{1}-\rme^{-2wa}\mathbf{R}_2\mathbf{R}_1\right)^{-1}\rme^{-2wa}\mathbf{R}_2\mathbf{T}_m\boldsymbol{\mathcal{G}_{m+}}.
\end{eqnarray}
The exponential factors in (\ref{Gm1s}) are deduced from the propagation distances of the waves, as described in Section~\ref{Green} . 

It remains to find the operators $\mathbf{R}_m$, $\mathbf{T}_m$, and $\mathbf{T}_1$. This is done as for the operators $\mathbf{R}_1$ and $\mathbf{R}_2$ in Section~\ref{Green} . We require the unit vectors in the directions of the $\mathbf{E}$- and $\mathbf{B}$-polarizations in the plate; for the $\mathbf{E}$-polarization the direction is
\[
\mathbf{n}_{Em}=\frac{1}{\sqrt{u^2+v^2}}\left(\begin{array}{c} 0 \\ -v \\ u \end{array}\right), \]
whereas for the $\mathbf{B}$-polarization we have
\[ \mathbf{n}^{\pm}_{Bm}=\frac{1}{\kappa\sqrt{\varepsilon_1\mu_1}\sqrt{u^2+v^2}}\left(\begin{array}{c} \rmi (u^2+v^2) \\ \pm w_1u \\ \pm w_1v \end{array}\right),
\]
where $\mathbf{n}^{+}_{Bm}$ ($\mathbf{n}^{-}_{Bm}$) is for a right-moving (left-moving) plane wave. We can now write the operators $\mathbf{R}_m$, $\mathbf{T}_m$, and $\mathbf{T}_1$ in terms of scalar reflection and transmission coefficients:
\begin{eqnarray}
\mathbf{R}_m=r_{Em}\mathbf{n}_{Em}\otimes\mathbf{n}_{Em}+r_{Bm}(\mathbf{R}_x\mathbf{n}^+_{Bm})\otimes\mathbf{n}^+_{Bm},\label{opss1} \\
\mathbf{T}_m=t_{Em}\mathbf{n}_{Em}\otimes\mathbf{n}_{Em}-t_{Bm}(\mathbf{R}_x\mathbf{n}_{B1})\otimes\mathbf{n}^+_{Bm}, \\
\mathbf{T}_1=t_{E1}\mathbf{n}_{E1}\otimes\mathbf{n}_{E1}+t_{B1}\mathbf{n}^-_{Bm}\otimes\mathbf{n}_{B1}. \label{opss2}
\end{eqnarray}
The reflection and transmission coefficients in (\ref{opss1})--(\ref{opss2}) for the two polarizations are a standard result~\cite{jac}:
\begin{eqnarray}
r_{Em}=\frac{w_1-\mu_1w}{w_1+\mu_1w}=-r_{E1}, \qquad r_{Bm}=-\frac{w_1-\varepsilon_1w}{w_1+\varepsilon_1w}=-r_{B1}, \\
\fl   t_{Em}=\frac{2w_1}{w_1+\mu_1w}=1-r_{E1}, \qquad t_{Bm}=\frac{2\sqrt{\varepsilon_1\mu_1}w_1}{\mu_1w_1+\varepsilon_1\mu_1w}=\sqrt{\frac{\varepsilon_1}{\mu_1}}(1+r_{B1}), \\
\fl   t_{E1}=\frac{2\mu_1w}{w_1+\mu_1w}=1+r_{E1}, \qquad t_{B1}=\frac{2\sqrt{\varepsilon_1\mu_1}w}{w_1+\varepsilon_1w}=\sqrt{\frac{\mu_1}{\varepsilon_1}}(1-r_{B1}).
\end{eqnarray}
The Green tensor (\ref{Gm1s}) is now completely specified.

The quantity $(\mathds{1}-\rme^{-2wa}\mathbf{R}_2\mathbf{R}_1)^{-1}\mathbf{R}_2$ in (\ref{Gm1s}) has been expanded in~\ref{A1}, so we can immediately write the expanded form of the Green tensor:
\begin{equation} \label{Gm2s}
\mathbf{\widetilde{G}}_m=\mathbf{\widetilde{G}}_{mb}+\rme^{w_1(x'+x)}\left[\mathbf{R}_m+\rme^{-2wa}\mathbf{T}_1\mathbf{R}_x\mathbf{N}_2\mathbf{T}_m\right]\boldsymbol{\mathcal{G}_{m+}}.
\end{equation}
This is a suitable form for evaluation of the electromagnetic stress tensor. In macroscopic electromagnetism the stress tensor is~\footnote{It is shown in~\cite{dzy61} that the Casimir stress in a material is given by (\ref{stressms}); this result was challenged in~\cite{raa}, but see~\cite{comments}.}
\begin{equation} \label{stressms}
\bm{\sigma}_m=\varepsilon_0\langle \mathbf{\hat{D}}\otimes\mathbf{\hat{E}}\rangle+\mu_0^{-1}\langle \mathbf{\hat{H}}\otimes\mathbf{\hat{B}}\rangle
-\frac{1}{2}\mathds{1}(\varepsilon_0\langle \mathbf{\hat{D}}\cdot\mathbf{\hat{E}}\rangle+\mu_0^{-1}\langle \mathbf{\hat{H}}\cdot\mathbf{\hat{B}}\rangle).
\end{equation}
This is computed from
\begin{eqnarray}
\langle\mathbf{\hat{D}}(\mathbf{r})\!\otimes\!\mathbf{\hat{E}}(\mathbf{r})\rangle=-\frac{\hbar\mu_0}{\pi}\int_0^\infty \rmd\xi\,\varepsilon(\rmi \xi)\xi^2\mathbf{G}^S_m(\mathbf{r},\mathbf{r},\rmi \xi), \label{EEms} \\
\langle\mathbf{\hat{H}}(\mathbf{r})\!\otimes\!\mathbf{\hat{B}}(\mathbf{r})\rangle
=\frac{\hbar\mu_0}{\pi}\lim_{\mathbf{r'}\rightarrow\mathbf{r}}\int_0^\infty \rmd\xi\,\frac{1}{\mu(\rmi \xi)}\nabla\times\mathbf{G}^S_m(\mathbf{r},\mathbf{r'},\rmi \xi)\times\stackrel{\leftarrow}{\nabla'}, \label{BBms}
\end{eqnarray}
which follow from (\ref{EE}) and (\ref{BB}) and the standard definition of the $\mathbf{D}$ and $\mathbf{H}$ fields. The regularization procedure is again to subtract the diverging stress in the absence of the material boundary, which in this case means dropping the bare part $\mathbf{G}_{mb}$ of the material Green tensor~\cite{dzy61}. The stress tensor (\ref{stressms}) is found to be diagonal and has zero $xx$-component, as discussed in Section~\ref{Stress}.


\begin{thebibliography}{99}
\bibitem{lif55}
Lifshitz E M 1955 {\it Zh.\ Eksp.\ Teor.\ Fiz.}\ {\bf 29}, 94. [1956 {\it Soviet Physics, JETP} {\bf 2}, 73]
\bibitem{LL}
Landau L D, Lifshitz E M and Pitaevskii L P 1980 {\it Statistical Physics, Part 2} (Oxford: Butterworth-Heinemann)
\bibitem{cas48}
Casimir H B C 1948 {\it Proc.\ Kon.\ Ned.\ Akad.\ Wetenschap.}\ B {\bf 51} 793
\bibitem{teo78}
Teodorovitch E V 1978 {\it Proc.\ R.\ Soc.}\ A {\bf 362} 71
\bibitem{sch81}
Schaich W L and Harris J 1981 {\it J.\ Phys.\ F: Met.\ Phys.}\ {\bf 11} 65
\bibitem{lev89}
Levitov L S 1989 {\it Europhys.\ Lett.}\ {\bf 8} 499
\bibitem{pol90}
Polevoi V G 1990 {\it Zh.\ Eksp.\ Teor.\ Fiz.}\ {\bf 98} 1990  [1990 {\it Soviet Physics, JETP} {\bf 71} 1119].
\bibitem{mkr95}
Mkrtchian V E 1995 {\it Phys.\ Lett.}\ A {\bf 207} 299
\bibitem{bar96}
Barton G 1996 {\it Ann.\ Phys.}\ {\bf 245} 361
\bibitem{pen97}
Pendry J B 1997 {\it J.\ Phys.: Condens. Matter} {\bf 9} 10301
\bibitem{per98}
Persson B N J and Zhang Z 1998 {\it Phys.\ Rev.}\ B {\bf 57} 7327
\bibitem{vol99}
Volokitin A I and Persson B N J 1999 {\it J.\ Phys.: Condens. Matter} {\bf 11} 345; Volokitin A I and Persson B N J 2002 {\it Phys.\ Rev.}\ B {\bf 65} 115419; Volokitin A I and Persson B N J 2006 {\it Phys.\ Rev.}\ B {\bf 74} 205413
\bibitem{kar99}
Kardar M and Golestanian R 1999 {\it Rev.\ Mod.\ Phys.}\ {\bf 71} 1233
\bibitem{kri02}
Krim J 2002 {\it Am.\ J.\ Phys.}\ {\bf 70}, 890
\bibitem{dav05}
Davies P C W 2005 {\it J.\ Opt.\ B: Quantum Semiclass.\ Opt.}\ {\bf 7} S40
\bibitem{dzy61}
Dzyaloshinskii, I E, Lifshitz E M and Pitaevskii L P 1961 {\it Adv.\ Phys.}\ {\bf 10} 165
\bibitem{nature}
Ball P 2007 {\it Nature} {\bf 447} 772
\bibitem{cap07}
Capasso F, Munday J N, Iannuzzi D and Chan H B 2007 {\it IEEE J. 
Select. Top. Quantum Electr.} {\bf 13} 400
\bibitem{mil94}
Milonni P W 1994 {\it The Quantum Vacuum} (London: Academic Press).
\bibitem{bir84}
Birrell N D and Davies P C W 1984 {\it Quantum fields in curved space}
(Cambridge: Cambridge University Press)
\bibitem{LLcm}
Landau L D, Lifshitz E M and Pitaevskii L P 1984 {\it Electrodynamics of Continuous Media,} 2nd ed.\ (Oxford: Butterworth-Heinemann)
\bibitem{GREE}
Leonhardt U and Philbin T G 2006 {\it New J. Phys.} {\bf 8} 247; Leonhardt U and Philbin T G 2008 {\it Progress in Optics} (in press) (arXiv:0805.4778v2 [physics.optics])
\bibitem{bar78}
Barash Y S 1978 {\it Izv.\ Vyssh.\ Uchebn.\ Zaved., Radiofiz.} {\bf 12} 1637 [1978 {\it Radiophysics and Quantum Electronics} {\bf 21} 1138]
\bibitem{mun05}
Munday J N, Iannuzzi D, Barash Y and Capasso F 2005 {\it Phys.\ Rev.}\ A {\bf 71} 042102 Erratum 2008 {\it Phys.\ Rev.}\ A {\bf 78} 029906(E)
 \bibitem{phi08}
Philbin T G and Leonhardt U 2008 {\it Phys.\ Rev.}\ A {\bf 78} 042107
 \bibitem{ros08}
Rosa F S S, Dalvit D A R and Milonni P W 2008 {\it Phys. Rev.}\ A {\bf 78} 032117
\bibitem{jac}
Jackson J D 1999 {\it Classical Electrodynamics} 3rd ed.\ (New York: Wiley)
\bibitem{leo}
Leonhardt U {\it Essential Quantum Optics: From Quantum Measurements To Black Holes} (Cambridge: Cambridge University Press, in press)
\bibitem{lou}
Loudon R 2000 {\it The Quantum Theory of Light} 3rd ed.\ (Oxford: Oxford University Press).
\bibitem{abs}
Kn\"{o}ll L, Scheel S and Welsch D-G 2001 QED in dispersing and absorbing media {\it Coherence and Statistics of Photons and Atoms} ed J Perina (New York: Wiley)
\bibitem{moc06}
Moch\'{a}n W L and Villarreal C 2006 {\it New J. Phys.} {\bf 8} 242
\bibitem{sch78}
Schwinger J, DeRaad L L and Milton K A 1978 {\it Ann.\ Phys.}\ {\bf 115} 1
\bibitem{raa}
Raabe C and Welsch D-G 2005 {\it Phys.\ Rev.} A {\bf 71} 013814
\bibitem{comments}
Pitaevskii L P 2005 {\it Phys.\ Rev.} A {\bf 73} 047801; Brevik I and Ellingsen S A, arXiv:0806.2927v1 [quant-ph]
\bibitem{vol08}
Volokitin A I and Persson B N J 2008 {\it Phys.\ Rev.}\ B {\bf 78} 155437

\end{thebibliography}
\end{document}